     \documentstyle[12pt,my,epsfig]{article}   
 
     \newlength{\dinwidth}                       
     \newlength{\dinmargin}                      
\def\lsim{\mathrel{\rlap{\lower4pt\hbox{\hskip1pt$\sim$}}
    \raise1pt\hbox{$<$}}}                % less than or approx. symbol
\def\gsim{\mathrel{\rlap{\lower4pt\hbox{\hskip1pt$\sim$}}
    \raise1pt\hbox{$>$}}}                % greater than or approx. symbol

\newcommand{\tqb}{P_{max}}
\newcommand{\kt}{k_{t}}
\newcommand{\ktp}{k_{t}^{\prime}}
\newcommand{\SMALLXC}{SMALLXa,SMALLXb}
\newcommand{\CCFM}{CCFMa,CCFMb,CCFMc,CCFMd}
\newcommand{\BFKL}{BFKLa,BFKLb,BFKLc}
\newcommand{\LDC}{LDCa,LDCb,LDCc,LDCd}
\newcommand{\alphasb}{\bar{\alpha}_s}
\newcommand{\JETSET}{Jetseta,Jetsetb,Jetsetc}
\newcommand{\LEPTO}{Ingelman_LEPTO65}
\newcommand{\PYTHIA}{Jetsetc}

\def\CASCADE{{\sc Cascade}}
\def\SMMOD{{\sc Smmod}}
\def\SMALLX{{\sc Smallx}}
\begin{document}

\input feynman 
\bigphotons 
\vspace*{10mm}
\begin{center}  \begin{Large} \begin{bf}
CCFM prediction on forward jets and $F_2$:\\
parton level predictions and a new 
hadron level Monte Carlo generator \CASCADE\\
%forward evolution parton level program \SMALLX~ and a new backward evolution
%hadron level Monte Carlo generator \CASCADE\\
%Initial state cascade effects according to CCFM and a backward evolution scheme 
%implemented in a hadron level Monte Carlo\\
  \end{bf}  \end{Large}
  \vspace*{5mm}
  \begin{large}
 H. Jung\\
  \end{large}
Physics Department, Lund University, Box 118, S-221~00 Lund, Sweden\\  
\end{center}
\begin{quotation}
\noindent
{\bf Abstract:}
A solution of the CCFM equation for a description of both the 
structure function $F_2$ and the cross
section of forward jet production as measured by the HERA experiments
 is obtained on the basis of the parton level  Monte Carlo
program \SMALLX. The treatment of the non - Sudakov form factor and the so -
called ``consistency constraint" are discussed.
Following this a backward evolution scheme according to CCFM is developed which
then is used to construct an efficient hadron level Monte Carlo program
\CASCADE.
The results from the forward evolution and the backward evolution Monte Carlos
are compared and found to be consistent. 
\end{quotation}
\section{Introduction}
The parton evolution at small values of $x$ is believed to be best described by
the CCFM evolution equation~\cite{\CCFM}, which for $x \to 0$ is equivalent to
the BFKL evolution equation~\cite{\BFKL} and for large $x$ reproduce the
standard DGLAP equations. The CCFM evolution equation takes coherence effects
of the radiated gluons into account via angular ordering.
\par
Already in 1992 the CCFM evolution equation was implemented in the
 parton level Monte Carlo program, \SMALLX~\cite{\SMALLXC},
using a forward evolution scheme.
In 1997 the Linked Dipole Chain~\cite{\LDC} and the corresponding
LDC hadron level Monte Carlo program  were 
developed based on a reformulation of the original CCFM equation.  
Predictions of the CCFM equation for hadronic final state properties were
studied in~\cite{Salam}, 
paying special attention to non-leading effects. 
Common to all three approaches was the difficulty to describe the structure
function $F_2$ and the cross section of forward jet production in
deep inelastic scattering at the same time~\cite{\LDC,Salam,Turnau}.
\par
This article is divided into two parts: 
In the first part I shall
describe briefly the implementation of the CCFM
evolution equation into the program \SMALLX~\cite{\SMALLXC},
 and discuss the
treatment of the non-Sudakov form factor 
$\Delta_{ns}$ as well as the effects of
the so-called ``consistency constraint", which was found to be 
necessary to include
non-leading contributions to the BFKL equation~\cite{Martin}. 
Then I demonstrate 
that a proper treatment of the non-Sudakov form factor 
is essential to achieve a
good description of $F_2$ and the forward jet data at the same time.
\\
In the second part 
 I describe a model for a backward evolution scheme
according to the CCFM equation, which forms the basis of the new hadron level
Monte Carlo generator \CASCADE. One ingredient in this 
approach is the determination of
the unintegrated gluon density function, which is obtained from the forward
evolution in \SMALLX. 
As a check of consistency 
I compare results obtained from \SMALLX~ with those from \CASCADE~and show nice
agreement.

\section{Forward evolution: CCFM and \SMALLX}
The implementation of the CCFM~\cite{\CCFM} parton evolution in the forward
evolution Monte Carlo program \SMALLX~ is described in detail in~\cite{\SMALLXC}. 
Here I only concentrate on the basic ideas and discuss the treatment of the
non-Sudakov form factor.
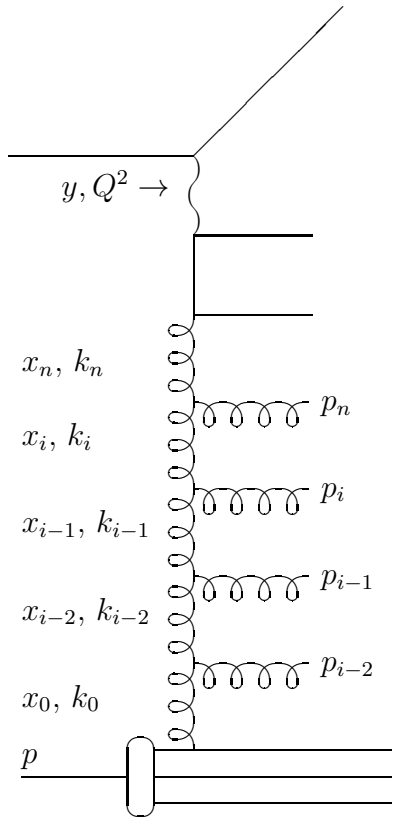
\begin{figure}
%%%%%%%%%%%%%%%%%%%%%%%%%%%%%%%%%%%%%%%%%%%%%%%%%%%% 
% begin picture kinematic variables 
%%%%%%%%%%%%%%%%%%%%%%%%%%%%%%%%%%%%%%%%%%%%%%%%%%%% 
\begin{center} 
\begin{picture}(30000,25000) 
% draw photon line 
\drawline\fermion[\NE\REG](5000,25000)[8000] 
\drawline\fermion[\E\REG](-2000,25000)[7000] 
\drawline\photon[\S\REG](5000,25000)[3] 
\global\advance\pmidx by -5000 
\put(\pmidx,\pmidy) {$y,Q^2 \to$ } 
\drawline\fermion[\E\REG](\photonbackx,\photonbacky)[4500] 
\global\advance\pmidy by 400 
\global\advance\fermionbackx by +1000 
\global\advance\pbackx by 500 
\global\advance\pbacky by -3200 
\global\advance\Yfive by + 3800 
%draw propagator 
\drawline\fermion[\S\REG](\photonbackx,\photonbacky)[3000] 
\global\advance\pmidx by -4000 
%\put(\pmidx,\pmidy) {$\xbj,Q^2 \to$ } 
%\drawline\gluon[\E\REG](\fermionbackx,\fermionbacky)[2] 
\global\Xseven = \pbackx 
\global\Yseven = \pbacky 
\global\advance\Xseven by + 500 
\global\advance\Yseven by - 300 
%\put(\Xseven,\Yseven){$q_{T\;i}$} 
\global\Xone = \pbackx 
\global\Yone = \pbacky 
\global\advance\Xone by + 4500 
\global\advance\Yone by - 750 
%\drawline\fermion[\S\REG](\fermionbackx,\fermionbacky)[1500] 
 
%\drawline\gluon[\E\REG](\fermionbackx,\fermionbacky)[3] 
\global\Xeight = \pbackx 
\global\Yeight = \pbacky 
\global\advance\Xeight by + 500 
\global\advance\Yeight by - 300 
%\put(\Xeight,\Yeight){$q_{T\;i+1}$} 
 
%\drawline\fermion[\S\REG](\fermionbackx,\fermionbacky)[2500] 
\global\advance\pmidx by -5000 
%\put(\pmidx,\pmidy) {$x_{\gamma},\mu^2 \to $} 
%draw q line 
\drawline\fermion[\E\REG](\fermionbackx,\fermionbacky)[4500] 
\global\Xeight = \pbackx 
\global\Yeight = \pbacky 
\global\advance\Xeight by + 500 
\global\advance\Yeight by - 300 
%\put(\Xeight,\Yeight){$q_{i}$} 
\global\advance\pmidy by 400 
%\put(\fermionbackx,\pmidy){$q$} 
\global\advance\pbackx by 500 
\global\advance\pbacky by -1500 
\global\Xsix = \pbackx 
\global\Ysix = \pbacky 
\global\advance\Ysix by + 2800 
%draw g line 
\drawline\gluon[\S\REG](\fermionfrontx,\fermionfronty)[3] 
\global\Xtwo = \pmidx 
\global\Ytwo = \pmidy 
\global\advance\Ytwo by - 500 
\global\advance\Xtwo by - 6500 
\put(\Xtwo,\Ytwo){{$x_n$, $k_n  $}} 
%\put(\Xthree,\Ythree){ hard scattering } 
\drawline\gluon[\E\REG](\gluonbackx,\gluonbacky)[4] 
\global\Xeight = \pbackx 
\global\Yeight = \pbacky 
\global\advance\Xeight by + 500 
\global\advance\Yeight by - 300 
\put(\Xeight,\Yeight){$p_n$} 
\global\advance\pmidy by 400 
%\put(\gluonbackx,\pmidy){$g$} 
\global\advance\pbacky by +400 
%\put(\pbackx,\pbacky){$x_j \gg \xbj$} 
\global\advance\pbacky by -1500 
%\put(\pbackx,\pbacky){$p^2 _t \sim Q^2$} 
\drawline\gluon[\S\REG](\gluonfrontx,\gluonfronty)[3] 
\global\advance\pmidx by -3000 
%\put(\pmidx,\pmidy) {$g \to $} 
\put(\Xtwo,\pmidy){{$x_i$, $k_{i}  $}} 
\drawline\gluon[\E\REG](\gluonbackx,\gluonbacky)[4] 
\global\Xeight = \pbackx 
\global\Yeight= \pbacky 
\global\advance\Xeight by + 500 
\global\advance\Yeight by - 300 
\put(\Xeight,\Yeight){$p_{i}$} 
\drawline\gluon[\S\REG](\gluonfrontx,\gluonfronty)[3] 
\put(\Xtwo,\pmidy){{$x_{i-1}$, $k_{i-1}  $}} 
\drawline\gluon[\E\REG](\gluonbackx,\gluonbacky)[4] 
\global\Xeight = \pbackx 
\global\Yeight= \pbacky 
\global\advance\Xeight by + 500 
\global\advance\Yeight by - 300 
\put(\Xeight,\Yeight){$p_{i-1}$} 
\drawline\gluon[\S\REG](\gluonfrontx,\gluonfronty)[3] 
\put(\Xtwo,\pmidy){{$x_{i-2}$, $k_{i-2}  $}} 
\drawline\gluon[\E\REG](\gluonbackx,\gluonbacky)[4] 
\global\Xeight = \pbackx 
\global\Yeight= \pbacky 
\global\advance\Xeight by + 500 
\global\advance\Yeight by - 300 
\put(\Xeight,\Yeight){$p_{i-2}$} 
\global\Xfour = \pbackx 
\global\Yfour = \pbacky 
\global\advance\Xfour by + 2500 
\global\advance\Yfour by + 2000 
 
\drawline\gluon[\S\REG](\gluonfrontx,\gluonfronty)[3] 
 \put(\Xtwo,\pmidy){{$x_{0}$, $k_{0}  $}} 

\global\advance\gluonbackx by -1500 
 
\multiput(\gluonbackx,\gluonbacky)(0,-1000){3}{\line(1,0){9000}} 
\global\advance\gluonbacky by -1000 
\global\advance\gluonbackx by -500 
\put(\gluonbackx,\gluonbacky){\oval(1000,3000)} 
\global\advance\gluonbackx by +2000 
\global\advance\gluonbacky by +1000 
%\put(\gluonbackx,\gluonbacky){$spectators$} 
\global\advance\gluonbackx by -2000 
\global\advance\gluonbacky by -1000 
\global\advance\gluonbackx by -500 
%draw initial proton 
\drawline\fermion[\W\REG](\gluonbackx,\gluonbacky)[4000] 
\global\advance\fermionbacky by -3000 
\global\advance\fermionbackx by 4000 
%\put(\fermionbackx,\fermionbacky){$ (b) $} 
\global\advance\fermionbacky by 3000 
\global\advance\fermionbackx by -4000 
%\global\advance\pbackx by 500 
\global\advance\pmidy by 500 
\put(\pbackx,\pmidy){$p$} 
\end{picture} 
\end{center} 
\caption{Kinematic variables for multi-gluon emission. The $t$-channel gluon
four - vectors are given by $k_i$ and the gluons emitted in the initial state
cascade have four - vectors $p_i$. 
\label{CCFM_variables} } 
\end{figure}
\par
The initial state gluon radiation is sketched in Fig.~\ref{CCFM_variables}
 together with a
definition of some variables used in the following.
According to the CCFM evolution equation, the 
emission of partons during the initial cascade is only allowed in  an angular
ordered phase space region. The maximum allowed angle $\Xi$ 
is defined by the hard
scattering quark box, which connects the gluon to the virtual photon. In 
terms of Sudakov variables the quark pair momentum is written as:
\begin{equation}
p_q + p_{\bar{q}} = Y (p_p + \Xi p_e) + Q_t
\end{equation}
where $p_p$ and $p_e$ are the proton and electron momenta, respectively
and $Q_t$ is the transverse momentum of the quark pair.
Similarly, the momenta $p_i$ of the gluons 
emitted during the initial state
cascade are given by (here treated massless):
\begin{equation}
p_i = y_i (p_p + \xi_i p_e) + p_{ti} \;  , \;\; \xi_i=\frac{p_{ti}^2}{s
y_i^2},
\end{equation}
with $y_i = (1 - z_i) x_{i-1}$ and $x_i = z_i x_{i-1}$ and $s=(p_p+p_e)^2$ being
the total electron proton center of mass energy. The variable $\xi_i$ is
connected to the angle of the emitted gluon with respect to the incoming proton
and $x_i$ and $y_i$ are the momentum fractions of the exchanged and emitted
gluons, while $z_i$ is the momentum fraction in the branching $(i-1) \to i$
and $p_{ti}$ is the transverse momentum of the emitted gluon.
\par
The angular ordered region is then specified by:
\begin{equation}
\xi_0 < \xi_1< \cdots < \xi_n < \Xi
\end{equation}
which becomes:
\begin{equation}
z_{i-1} q_{ti-1} < q_{ti} 
\end{equation}
when using the rescaled transverse momenta $q_{ti}$ of the emitted
gluons defined by:
\begin{equation}
 q_{ti} = x_{i-1}\sqrt{s \xi_i} = \frac{p_{ti}}{1-z_i}
\end{equation}
\par
In \SMALLX, the initial state gluon cascade is generated 
in a forward evolution approach from a initial distribution of the
$k_t$ unintegrated gluon distribution according to:
\begin{equation}
x G_0(x,k_{t\;0}^2) = N \cdot (1-x)^4 \cdot \exp{\left(-k_{t\;0}^2/k_0^2\right)}
\label{eq1}
\end{equation}
where $N$ is a normalization constant.
The exponential factor in eq.(\ref{eq1}) is a gaussian distribution which
specifies the distribution of the initial transverse momenta $k_{t\;0}$. In the
following we set $k_0^2=1$~GeV$^2$, corresponding to a gaussian width of 
$\sigma = k_0/\sqrt{2} = 0.7$ GeV.
The input gluon distribution needs to be adjusted to fit existing data, but it
turns out that the small $x$ behavior of the structure function $F_2$
is rather
insensitive to the actual choice of $xG_0(x,k_t^2)$ and only the normalization
$N$ acts as a free parameter.
\par
The initial state branching of a gluon $k_{i}$
into another virtual ($t$-channel) gluon $k_{i+1}$
 and a final gluon $p_{i+1}$ (treated on mass shell) is 
  generated successively from the inital gluon
distribution at a starting scale $Q_0$. 
The probability for successive branchings to occur is given by the 
CCFM splitting function~\cite{\CCFM}:
\begin{equation}
d{P}_i  = \tilde{P}_g^i(z_i,q^2_{i},k^2_{ti}) \cdot \Delta_s d z_i
             \frac{d^2 q_{i} }{\pi q^{2}_{i}} 
	   	 \cdot \Theta(q_{i}-z_iq_{i-1})
		 \cdot \Theta(1-z_i-\epsilon_i)
		 \label{CCFM_splitting}
\end{equation}
with $q_{i}=p_{ti}/(1-z_i)$ being the rescaled transverse momentum
of the emitted gluon $i$. The fractional energy of the exchanged gluon $i$ is
given by $x_i$ and the energy transfer between the exchanged gluon $i-1$
and $i$ is given by
$z_i=x_i/x_{i-1}$. A collinear cutoff 
 $\epsilon_i=Q_0/q_{i}$ is introduced  to avoid
the $1/(1-z)$ singularity. The Sudakov form factor $\Delta_s$  is given by: 
\begin{equation}
\Delta_s(q_{i},z_iq_{i-1}) =\exp{\left(
 - \int_{(z_{i-1} q_{i-1})^2} ^{q^{2}_{i}}
 \frac{d q^{2}}{q^{2}} 
 \int_0^{1-Q_0/q} dz \frac{\alphasb(q^2(1-z)^2)}{1-z}
  \right)}
  \label{Sudakov}
 \end{equation}
 with $\alphasb=\frac{C_A \alpha_s}{\pi}=\frac{3 \alpha_s}{\pi}$. The Sudakov
  form factor cancels
 at an inclusive level against the $1/(1-z)$ collinear singularity
 of the splitting function.
 Coherence effects are taken into account
by angular ordering 
$q_{i} > z_{i-1} q_{i-1}$
given by the first $\Theta$ function in 
eq.(\ref{CCFM_splitting}). The distribution of successive values of 
$q_{i}$ is controlled by this Sudakov form factor.
The cascade continues until $q_i$ reaches $\tqb$,
 which is given by the partons
from the hard scattering matrix element.
\par
The gluon splitting function $\tilde{P}_g^i$ is given by:
\begin{equation}
\tilde{P}_g^i= \frac{\alphasb(q^2_{i}(1-z_i)^2)}{1-z_i} + 
\frac{\alphasb(k^2_{ti})}{z_i} \Delta_{ns}(z_i,q^2_{i},k^2_{ti})
\label{Pgg}
\end{equation}
with $k_{ti}$ being the transverse momentum of the exchanged gluon $i$
and the non-Sudakov form factor $\Delta_{ns}$ being defined as:
\begin{equation}
\log\Delta_{ns} =  -\alphasb(k^2_{ti})
                  \int \frac{dz'}{z'} 
			\int \frac{d q^2}{q^2} 
              \Theta(k_{ti}-q)\Theta(q-z'q_{ti})
		  \label{non_sudakov}			
\end{equation}
The difference to the DGLAP splitting function for gluons is the appearance of
the non-Sudakov form factor 
$\Delta_{ns}$, which screens the $1/z$ singularity in eq.(\ref{Pgg}).
 The finite
terms in the splitting function are neglected, since they are not obtained in
CCFM at the leading 
infrared accuracy~\cite[p.72]{CCFMc}.
In the region $k^2_{ti} > z_i q^2_{i}$, 
$\Delta_{ns}$ can be written as:
\begin{equation}
\log\Delta_{ns} = - \alphasb(k^2_{ti})
\log\left(\frac{1}{z_i}\right)\log\left(\frac{k^2_{ti}}{z_i q^2_{i}}\right)
\label{ns_simple}
\end{equation}
where the limits of the $z^{\prime}$ integral 
in eq.(\ref{non_sudakov})
were set to $z_i < z^{\prime} <
1$. However, the $\Theta$ functions in eq.(\ref{non_sudakov})
 give further constraints on
the upper limit of the $z^{\prime}$ integral\footnote{I am very grateful to 
J. Kwiecinski for the explanation of these limits}: 
\begin{equation}
z_i \leq z^{\prime} < z_0=\mbox{max}\left(1,\frac{k_{ti}}{q_{ti}}\right)
\end{equation}
In the region $z_i > k_{ti}/q_{i}$, the $z^{\prime}$ integral becomes zero.
Putting all this together the 
non-Sudakov form factor can be expressed as~\cite{Martin_Sutton}:
\begin{equation}
\log\Delta_{ns} = -\frac{ \alpha_s(k^2_{ti})}{2 \pi }
\log\left(\frac{z_0}{z_i}\right)
\log\left(\frac{k^2_{ti}}{z_0z_i q^2_{i}}\right)
\label{ns_new}
\end{equation} 
where
$$z_0 = \left\{ \begin{array}{ll}
              1             & \mbox{if  } k_{ti}/q_{i} > 1 \\
		  k_{ti}/q_{i} & \mbox{if  } z_i < k_{ti}/q_{i} \leq 1 \\
		  z_i             & \mbox{if  } k_{ti}/q_{i} \leq z_i  
		  \end{array} \right. $$
which means that in the region $k_{ti}/q_{i} \leq z_i$ we have $\Delta_{ns}=1$,
giving no suppression at all.
\par
The use of eq.(\ref{ns_simple}) in the region
$k^2_{ti} \leq z_i q^{2}_{i}= z_i p_{t\;i}^2/(1-z_i)^2 $
 would result in a change of the sign, and
therefore instead of giving a suppression form factor would give an
 enhancement,
which is in contradiction to a form factor. The constraint\footnote{
The constraint: $Q^2_{i-1} < Q_{i}^2/z_i$ was already derived
in \protect\cite[eq.(2.11)]{BCM}
and in \protect\cite{PYTHIAPSb} in a frame where parton $i$ is along the $z$
axis. In \protect\cite{CCFMc} this was used to derive 
 a similar condition on the transverse momenta (see 
 \protect\cite[eq.(7.23)]{CCFMc}) $k_t^2 > z q^{2}$, when the
 transverse momenta are the dominant contribution to the virtualities.
 However, in a different frame (the lab frame used in CCFM for example),
 the numerical value of $z$ might be different, this constraint might 
 be violated.} 
$k^2_{ti} > z_i q^2_{i}$ is often referred to as the 
``consistency constraint"~\cite{Martin}.
\par
In the original version of \SMALLX~\cite{\SMALLXC}
 the splitting function was defined as:
\begin{equation}
\tilde{P}_g^i= \frac{\alphasb(q^2_{i}(1-z_i)^2)}{1-z_i} + 
\frac{\alphasb(k^2_{ti})}{z_i} \frac{\Delta_{ns}(z_i,q^2_{i},k^2_{ti})}
{\int dz'\Delta_{ns}(z',q^2_{i},k^2_{ti}) }
\end{equation}
which also is valid over the full phase space, but gives only very weak
suppression of the small $z$ splittings, and therefore overshoots the 
measured structure
function $F_2$ at small values of $x$.
\par
The Monte Carlo program \SMALLX~\cite{\SMALLXC} has been modified to include the
non-Sudakov form factor according to eq.(\ref{ns_new}) and $k_{t}^2$
was used as the scale in
$\alpha_s$ also in the 
 matrix element. To avoid problems at small $k_t^2$, 
a upper limit on $\alpha_s$ was introduced such that
 $\alpha_s(k^2_{ti}) \leq 0.6$. 
This modified version of \SMALLX~ is called 
 \SMMOD~  in the following.
\par
A few words are needed concerning the
``consistency constraint", which was introduced to
account for next-to-leading effects in the BFKL equation.
Including this constraint it was found in 
\cite{Martin}, that about 70\% of the full
next-to-leading order corrections to the BFKL equation are simulated. However,
in LO BFKL the true kinematics of the branchings are neglected. They are
included only in NLO 
therefore act
as next-to-leading order effects. 
Therefore this constraint is often also called
``kinematic constraint". In the CCFM equation
 energy and momentum conservation
is already included at LO by the full treatment of the radiated gluons,
 and it is not clear, whether the arguments coming
from BFKL also apply to CCFM. 
\par
\begin{figure}[htb]
  \vspace*{2mm}
\epsfig{figure=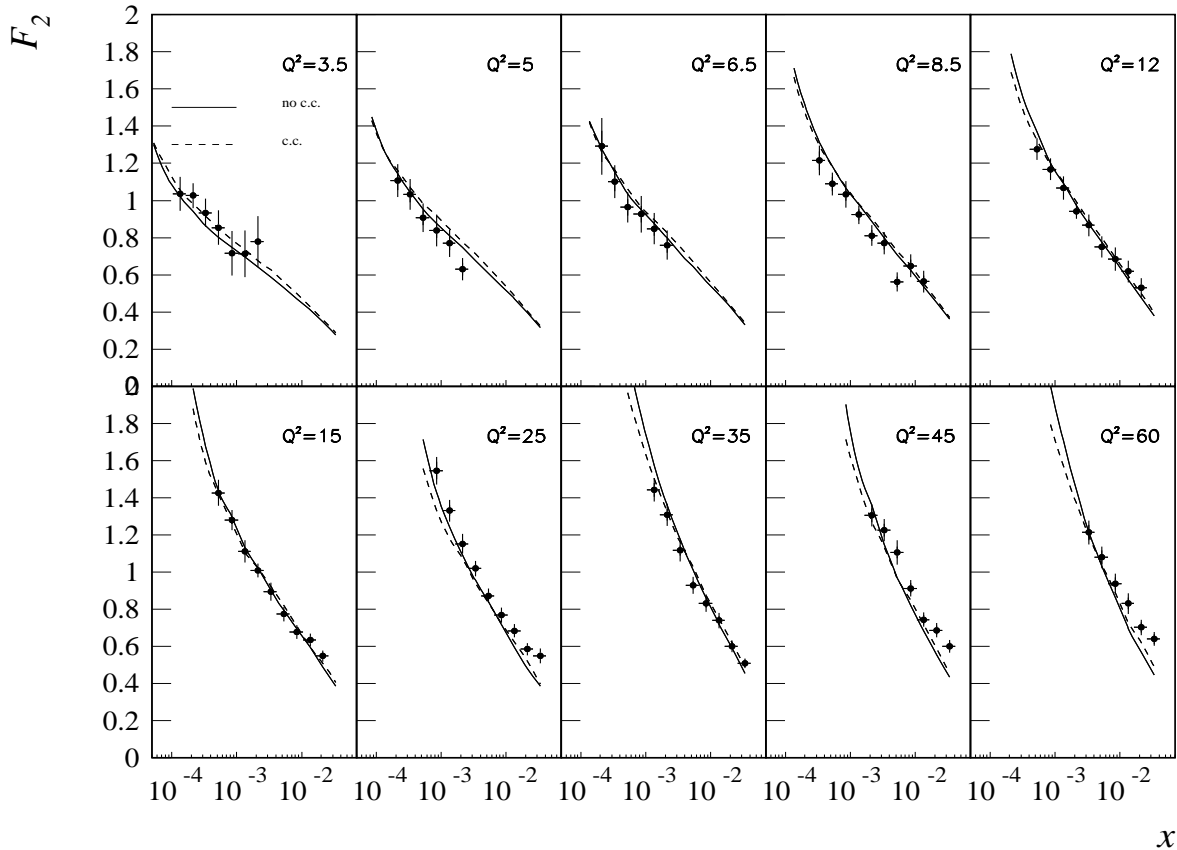,
width=17cm,height=12cm}
\caption{{\it
The structure function $F_2(x,Q^2)$ compared to 
H1 data~\protect\cite{H1_F2_1996}.
 The solid (dashed) line is 
the prediction of the \SMMOD~ Monte Carlo without (with)
applying the ``consistency constraint" (c.c.).
  }}\label{f2_smallx}
\end{figure}
In the following the effects of 
the ``consistency constraint" are studied in more detail.
In Fig.~\ref{f2_smallx} the structure function $F_2(x,Q^2)$ as calculated from
\SMMOD~ (including the non-Sudakov form factor according to eq.(\ref{ns_new}))
 is shown. The solid curve is  without the ``consistency constraint",
 the dashed curve shows the result after applying the
``consistency constraint". The collinear cutoff $Q_0$ was set to
$Q_0 =1.1 (0.85) $~GeV (without (with) ``consistency constraint", respectively).
The values of $Q_0$ have been chosen such, that a reasonable description of
$F_2$ is obtained.
\begin{figure}[htb]
  \vspace*{2mm}
\epsfig{figure=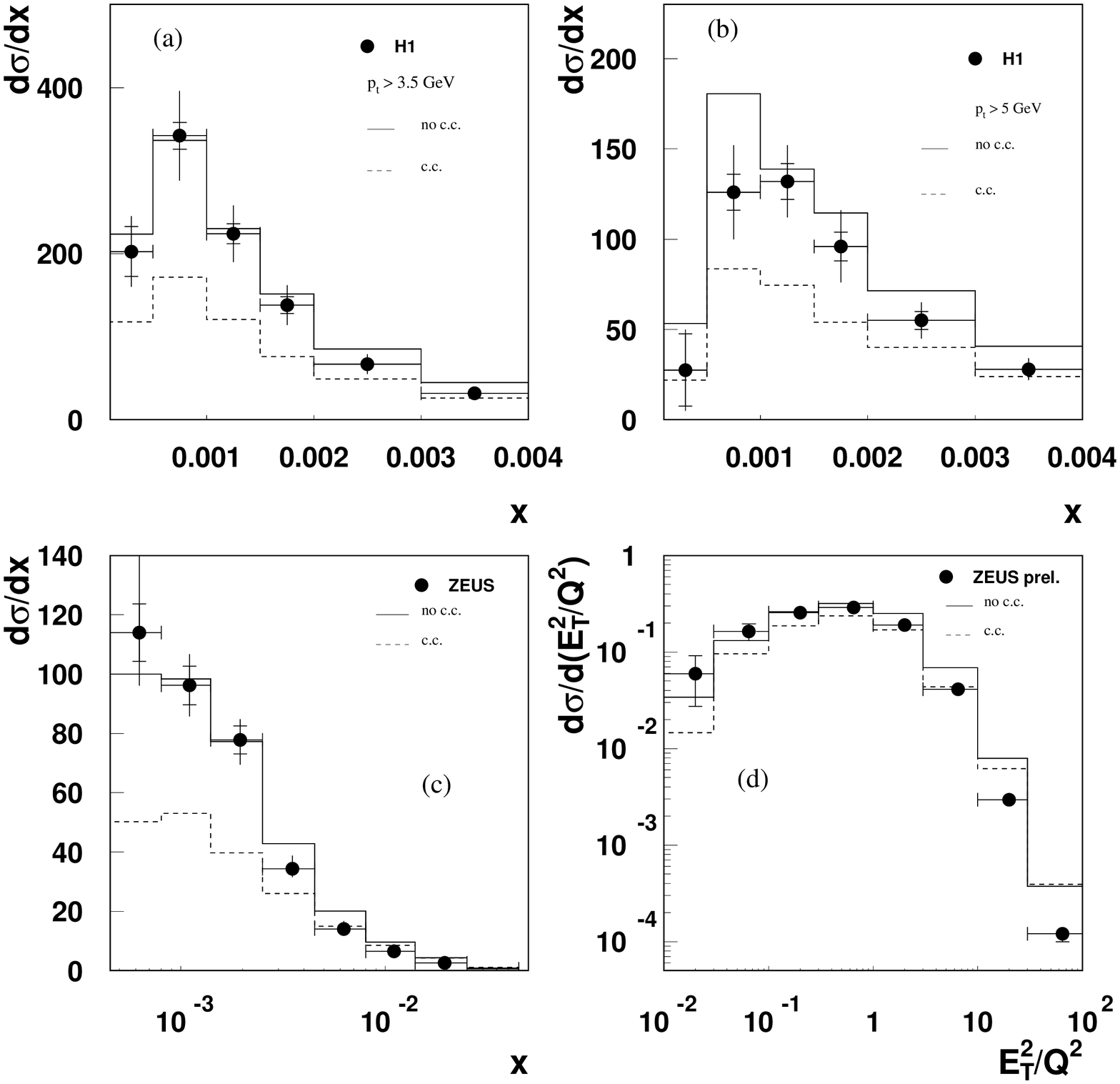,
width=17.5cm,height=17cm}
\caption{{\it
$a.-c.$ The cross section for forward jet production as a function of $x$, 
for different cuts in $p_t$ compared to H1 data~\protect\cite{H1_fjets_data}
($a.-b.$) and 
compared to ZEUS data~\protect\cite{ZEUS_fjets_data} ($c.$).
$d.$ The cross section for forward jet production as a function of $E^2_T/Q^2$
compared to \protect\cite{ZEUS_fjets_pt2/q2}.
 The solid (dashed) line is 
the prediction of the \SMMOD~ Monte Carlo without (with)
applying the ``consistency constraint" (c.c.).
  }}\label{fwdjet_smallx}
\end{figure}
In Fig.~\ref{fwdjet_smallx} the 
cross section of forward jet production
and the measurements of
 \cite{H1_fjets_data,ZEUS_fjets_data,ZEUS_fjets_pt2/q2} are shown
as a function of $x$ (Fig.~\ref{fwdjet_smallx}a.-c.)
 and as a function of
$E^2_T/Q^2$ (Fig.~\ref{fwdjet_smallx}d.),
 using the same parameter setting as for
the calculation of $F_2$. This plots have been produced with the HZTOOL 
analysis package~\cite{hztool}.
Nice agreement with the data is observed for the case
without the ``consistency constraint". When the  
``consistency constraint" is applied, the prediction falls below the data. One
has to note that in all cases the scale for $\alpha_s$ is $k_t^2$. Comparing
this  to
the results in~\cite{Martin} the same trend is found: applying the
``consistency constraint" and using $k_t^2$ as the scale in 
 $\alpha_s$, the prediction of the cross section for 
 forward jet production falls below
 the data, whereas there is essentially no effect seen in the prediction of the
 structure function $F_2$. It is interesting to note that the BFKL result 
 of \cite{Martin} (dashed curve in Fig.~4 of \cite{Martin}) seems to be
 identical to the dashed line in Fig.~\ref{fwdjet_smallx} obtained from \SMMOD. 

\section{Backward evolution: CCFM and \CASCADE}

The forward evolution procedure as implemented in \SMALLX~ is a direct way of
solving the CCFM evolution equation including the correct treatment of the
kinematics in each branching. However the forward evolution is rather
time consuming,
 since in each branching a weight factor is associated, and only 
after the initial state cascade has been generated completely,
 it can be decided whether
the kinematics allow to generate the hard scattering process. Quite often, a
complete event has to be rejected. 
\par
A more efficient procedure to be used in a full hadron level Monte Carlo is a
backward evolution scheme,
which is used in standard 
Monte Carlo programs~\cite{PYTHIAPSb,PYTHIAPSa} 
using a DGLAP type parton cascade. The idea is to 
first generate  
the hard scattering process with the initial parton momenta distributed
according to  the  parton distribution functions. 
This involves in general only a fixed
number of degrees of freedom, and the hard scattering process can be generated
quite efficiently. Then, the initial state cascade is generated by
going backwards
 from the hard scattering process towards the beam particles. In a DGLAP
type cascade the 
evolution (ordering) is done usually in the virtualities of the
exchanged $t$-channel partons.
\par
According to the CCFM equation
 the probability of finding a gluon in the proton depends
on three variables, the momentum fraction $x$, the transverse momentum squared
$k_t^2$ of the exchanged gluons and the maximum angle allowed for any emission
$\tqb = x_{n-1} \sqrt{s \Xi}$.
This probability distribution as a function of $x$, $k_t^2$ and $\tqb$
is obtained from a solution of the CCFM equation, here taken from the Monte
Carlo procedure used in \SMALLX~ and \SMMOD~ as described in the previous chapter.
For practical reasons 
the gluon density is obtained in a $50 \times 50 \times 20$ 
grid in $\log x$, $\log k_{t}$ and
$\log \tqb$, and then a linear interpolation method is used to 
obtain the gluon density at values in between the grid points.
\par
However there is a problem in the normalization of the gluon density: In \SMALLX~
and \SMMOD~ the normalization is done via the momentum sum rule for all gluons
below a given $k_t$ for all $\tqb$. But this $\tqb$ depends on the kinematics of
the process involved. In order to obtain a gluon distribution, which is
universal, the momentum sum rule is applied for each $\tqb$ separately.
Therefore
 differences between \SMMOD~ and the cross section calculated in the following
  are not surprising.
\par
To fully simulate the hadronic final state of a small $x$ process including the
CCFM parton evolution, a new hadron level Monte Carlo program, \CASCADE, has
been developed. It consists of three steps:
\begin{itemize}
\item[$\bullet$] 
First the hard scattering process is generated:
\begin{equation}
\sigma = \int dk_t^2 dx_g {\cal A}(x_g,k_t^2,\tqb)
 \sigma (\gamma^* g^* \to q \bar{q})
\label{x_section}
\end{equation}
where the off-shell matrix elements (given in~\cite[p. 178 ff]{off_shell_me})
 are used, with the gluon momentum (in Sudakov representation):
 \begin{equation}
 k = x_g p_p + \bar{x}_g p_e + k_t \sim x_g p_p  + k_t
 \end{equation}
The gluon virtuality $k^2$ is then $k^2 \sim k_t^2$. 
\item[$\bullet$]
The initial state cascade is generated according to CCFM in a backward evolution
approach (described in the next section). 
\item[$\bullet$]
The hadronization is performed using the Lund string fragmentation implemented
in JETSET \cite{\JETSET}.
\end{itemize}
\par
At the present state, final state gluon radiation from both the quarks and the
gluon $p_i$ are neglected.
\par
However in the backward evolution there is one difficulty: The gluon virtuality
enters in the hard scattering process and also influences the kinematics of the
produced quarks and therefore the maximum angle allowed for any further emission
in the initial state cascade. This virtuality is only known after the whole
cascade has been generated, since it depends on the history of the gluon
evolution.
 In the
evolution equations itself it does not enter, since there only the longitudinal
energy fractions $z_i$ and the transverse momenta are involved. 
This problem can only be approximatively overcome by using $k^2 = k_t^2/(1-x_g)$
for the virtuality
which is correct in the case of no further
gluon emission in the initial state. 
\par
 The
Monte Carlo program \CASCADE~ can be used to generate 
 unweighted full hadron level events, including initial state parton evolution
 according to the CCFM equation and the off - shell matrix elements for the hard
 scattering process. It is applicable both for photo production of heavy quarks
 as well as for deep inelastic scattering.  The typical time for
generating one event is $\sim 0.03 $ sec, which is similar to the time needed by
standard Monte Carlo event generators such as LEPTO~\cite{\LEPTO} or
PYTHIA~\cite{\PYTHIA}. 
 In standard Monte Carlo generators~\cite{\LEPTO,\PYTHIA},
  uncertainties concerning the
 scale to be used in the the structure functions and in $\alpha_s$, and also the
 upper virtuality used in the parton shower approach are typical.
 As in \SMMOD~ and \SMALLX, in \CASCADE, there are essentially
 no such free parameters left since everything is fixed 
 by the requirement to describe the
 inclusive structure function $F_2$.

\subsection{Backward evolution formalism}

The CCFM evolution equation can be written~\cite{CCFMd,Salam,Martin_Sutton}
 in a integral form as:
\begin{equation}
{\cal A} (x,\kt,\tqb ) = {\cal A}^0 (x,\kt,\tqb ) + \int \frac{dz }{z}
\int \frac{d^2 q}{\pi q^{2}} \Theta(\tqb - zq) \Delta_s(\tqb ,z q)
\tilde{P}(z,q,\kt) {\cal A}(\frac{x}{z},\ktp,q)
\label{CCFM_integral}
\end{equation} 
with $\ktp = | \vec{k}_{t} + (1-z) \vec{q}|$
 and with $\tqb$ being the upper scale for the last angle of the emission:
$\tqb > z_n q_n$, $q_n > z_{n-1} q_{n-1}$, ...,  
$q_{1} > z_0 q_{0}=Q_0$. Here $q$ is used as a shorthand notation for
2-dimensional vector of the rescaled transverse momentum 
$\vec{q}=\vec{q}_t=\vec{p}_t/(1-z)$. 
The splitting function
$\tilde{P}(z,q,\kt)$ is defined in eq.(\ref{Pgg}) and the Sudakov form factor
$\Delta_s(\tqb ,z q)$ is given in eq.(\ref{Sudakov}). 
The first term in eq.(\ref{CCFM_integral}) can be expressed as:
\begin{equation}
 {\cal A}^0 =  G_0(x,k_t^2) \Delta_s(\tqb ,k_0)
 \label{a0}
 \end{equation}
with $G_0(x,k_t^2)$ defined in eq.(\ref{eq1}) and $k_0$ being the starting
scale for the evolution.
 Thus ${\cal A}^0$ gives the
probability of moving from $x_0$, $k_{t0}$ and $k_0$ to $x$, $k_t$ and $\tqb$
without further branching. 
\par
In a backward evolution, we start from gluon $k_n$ 
(see Fig.~\ref{CCFM_variables}) and go successively down in the ladder until we
end up at gluon $k_0$. Therefore we need to reconstruct the branchings
$k_i \to k_{i-1} + q_i$, where $k_i$ refers to the
four vector of the $t$-channel gluon
closest to the hard interaction, $k_{i-1}$ to the one closer to the proton and
$q_i$ to the gluon emitted between $i$ and $i-1$.
The momenta of such a branching  with given 
$x_i$, $k_{ti}$ and $q_{i+1}$ are reconstructed 
as follows:\footnote{I am very much grateful 
to G. Salam for suggesting this approach for a
backward evolution.}
\begin{itemize}
\item[$a.$]
The value $\tilde{q}_{i}=z_{i}q_{i}$ is calculated from 
$d^2\tilde{q}_{i}/\tilde{q}_{i}^2 \Delta_s(q_{i+1},\tilde{q}_{i})$ 
with the Sudakov form factor defined in
eq.(\ref{Sudakov}).
 Note, that due to the angular ordering constraint only the
product $z_{i}q_{i}$ can be obtained, whereas in the DGLAP case with
transverse momentum ordering, $q_{i}$ could be obtained already here.
\item[$b.$]
Next the splitting variable $z_{i}=x_i/x_{i-1}$ needs to be generated.
 The correct
splitting function given in eq.(\ref{Pgg}) cannot be used, since
it involves the variable $q_{i}$ (in  the argument of $\alpha_s$), which
is not known at this stage. Thus,
$z_{i}$ is generated first according to a approximate splitting function
$P_{gg}^{appr}$:
\begin{equation}
P_{gg}^{appr}(z_{i},k_{t i}^2) = \frac{\alphasb(k_{ti})}{z_{i}} +
\frac{\alphasb(q_{t\,min})}{1-z_{i}} 
 \geq \tilde{P}_{gg}
\end{equation}
which is always larger than the true
splitting function. The limits on $z_{i}$ are given by:
$$ x_i \leq z_{i} \leq 1 - \frac{x_i Q_0}{q_{i+1}}.$$
Also the $z$ range needs to be larger than the true one:
$z_{i} < 1 - Q_0/q_{i}$, but $q_{i}$ is not determined at this stage.
Having $z_{i}$ generated, the true $q_{i}$ can be calculated with 
$q_{i}=\tilde{q}_{i}/z_{i}$. If now $z_{i} > 1 - Q_0/q_{i}$, the
branching is rejected, and a new one is generated by 
going back to step $a$.
\item[$c.$]
Now the transverse momentum of gluon $(i-1)$ can be calculated, by generating a
random angle $\phi$ around $q_{i}$: $k_{ti-1}^2 = k_{ti}^2 + q_{ti}^2 +
2 \sqrt{k_{ti}^2}\sqrt{q_{ti}^2}\cos{\phi}$.
\item[$d.$]
A  branching is accepted with the probability:
\begin{equation}
 \frac{\tilde{P}(z_{i},k_{t i}^2, q_{i}^2)}{P_{gg}^{appr}(z_{i},k_{t i}^2)} 
\frac{x_{i-1}{\cal A}(x_{i-1},k_{t i-1}^2, q_{i}^2)}
{x_{i}{\cal A}_{max}(x_{i})},
\label{prob_accept}
\end{equation}
where ${\cal A}_{max}(x_{i})$ is the maximum value, ${\cal A}$ can take for
any value $x>x_{i}$ and any value of $k_t^2$, $q_t^2$.
\end{itemize}
This then completes the reconstruction of one branching.
In the  next branching  generated the values of $x_{i-1}$, 
$k_{ti-1}$ and $q_{ti}$ play the role, the values $x_i$,$k_{ti}$ and $q_{ti+1}$
 had before. The cascade ends,
with the probability:
\begin{equation}
\frac{{\cal A}^0(x_{i-1},k_{t i-1}^2, q_{i}^2)}
{{\cal A}(x_{i-1},k_{t i-1}^2, q_{i}^2)} 
\end{equation}
This means that a further branching would have come from the initial gluon
distribution.
\par
Due to the complicated structure of the splitting function and  ${\cal A}$, this
procedure is quite inefficient. It can be considerably improved by a different
procedure in step $c$
in the selection of the $\phi$ angle and the transverse momentum 
$k_{t i-1}^2$: the angle $\phi$ is now selected according to the probability:
\begin{equation}
\frac{{\cal A}^0(x_{i-1},k_{t i-1}^2, q_{i}^2)}
{{\cal A}(x_{i-1},k_{t min}^2, q_{i}^2)} 
\end{equation}
where $k_{t min}^2 =k_{ti}^2 + q_{ti}^2 - 2 \sqrt{k_{ti}^2}\sqrt{q_{ti}^2}$ is
the minimum transverse momentum allowed in this branching. In step $d$ the
ratio of the parton densities (in eq.(\ref{prob_accept}))
 is then changed correspondingly to:
$$
\frac{x_{i-1}{\cal A}(x_{i-1},k_{t min}^2, q_{i}^2)}
{x_{i}{\cal A}_{max}(x_{i})}.
$$ 
This gives an improvement of a factor of $\sim 2$ in time.
\par
As mentioned already above, the true virtuality of the $t$-channel gluons can
only be reconstructed after the full cascade has been generated. Now by going
from the last (closest to the proton) gluon, which has virtuality
$k_0^2=k_{t\;0}/(1-x_0)$, forward in the cascade to the hard scattering process,
the true virtualities of $k_i^2$ are reconstructed.
At the end, the gluon entering to the quark box will have a larger virtuality
than without initial state cascade. Thus a check is performed, whether the
production of the quarks is still kinematically allowed. If not, then the whole
cascade is rejected, and the event without cascade is kept.

\section{Results}

In this section, results obtained from the 
 hadron level Monte Carlo program \CASCADE, described in the previous
chapter, are compared to those from \SMMOD. In all cases, no 
``consistency constraint" is applied. The unintegrated gluon distribution was
obtained from \SMMOD. At parton level, both programs are expected to be
identical, but differences can occur because of normalization problems,
the finite grid size used
to define of the unintegrated gluon density, and the effect
of the virtuality of the gluon entering the hard scattering process, which is
not properly known in the backward evolution approach.
\par
\begin{figure}[htb]
  \vspace*{2mm}
\epsfig{figure=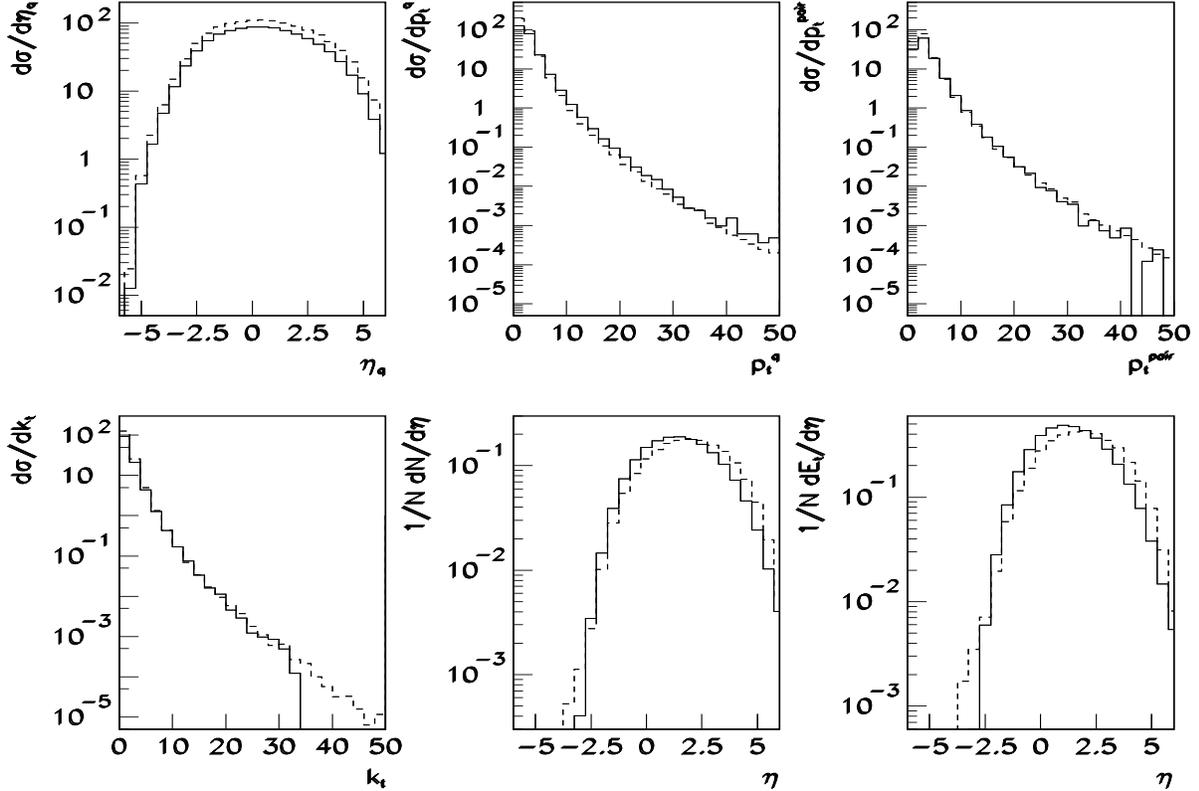,
width=17cm,height=12cm}
\caption{{\it
Comparison of the cross section obtained from the backward evolution 
Monte Carlo  \CASCADE~ (solid line)
with \SMMOD~ (dashed line) both at parton level only.
The upper 3 plots show the cross section is shown as a function of 
the quark rapidity $\eta_q$, the quark transverse momentum $p_t$ and the 
transverse momentum of the quark pair $p_t^{pair}$.
The lower 3 plots show the cross section as a function of the gluon transverse
momentum 
 $k_t$ and  the multiplicity and transverse energy flow 
 of the gluons from the initial
state cascade are shown as a function of the rapidity $\eta$
  }}\label{smallx_rinips}
\end{figure}
In Fig.~\ref{smallx_rinips} the cross section as a function of
the rapidity of the quarks $\eta_q$, 
the quark transverse momentum $p_t^q$, the 
transverse momentum of the quark pair $p_t^{pair}$ and 
the gluon transverse
momentum 
 $k_t$ obtained from \CASCADE~   are
compared to the ones obtained from  
\SMMOD~ (dashed line). A
rather good agreement is obtained.
Also shown is the 
multiplicity and the
transverse energy flow  as a function of rapidity for the gluons of the initial
state cascade, and the 
 prediction from \SMMOD~ for comparison. This  shows
that both the cross section and the initial state cascade are well reproduced
within the backward evolution approach. However, the differences seen are
due to uncertainties in 
the determination of the gluon density function. Further improvements
needed here to optimize the grid and the extraction of the gluon density.
\par
\begin{figure}[htb]
  \vspace*{2mm}
\epsfig{figure=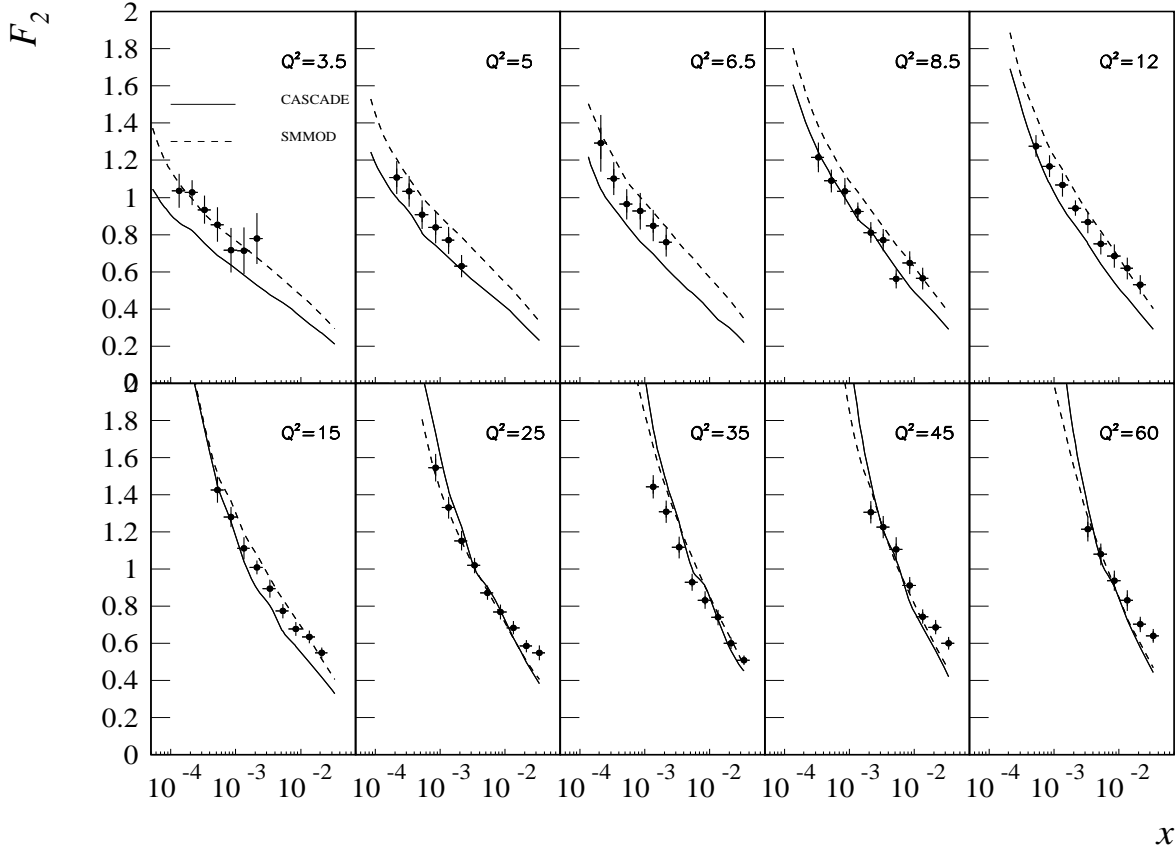,
width=17cm,height=12cm}
\caption{{\it
Comparison of the structure function $F_2$ 
 obtained from the backward evolution 
Monte Carlo  \CASCADE~(solid line)
with \SMMOD~ (dashed line).
  }}\label{f2comp}
\end{figure}
In Fig.~\ref{f2comp} the structure function $F_2$ obtained from \CASCADE~
 is compared to the one obtained from \SMMOD~ (the same as the solid
line in Fig.~\ref{f2_smallx}).
At large $Q^2$ the
results agree rather nicely, whereas at small $Q^2$ differences are seen, due to
uncertainties in the normalization of the gluon distribution.
\begin{figure}[htb]
  \vspace*{2mm}
\epsfig{figure=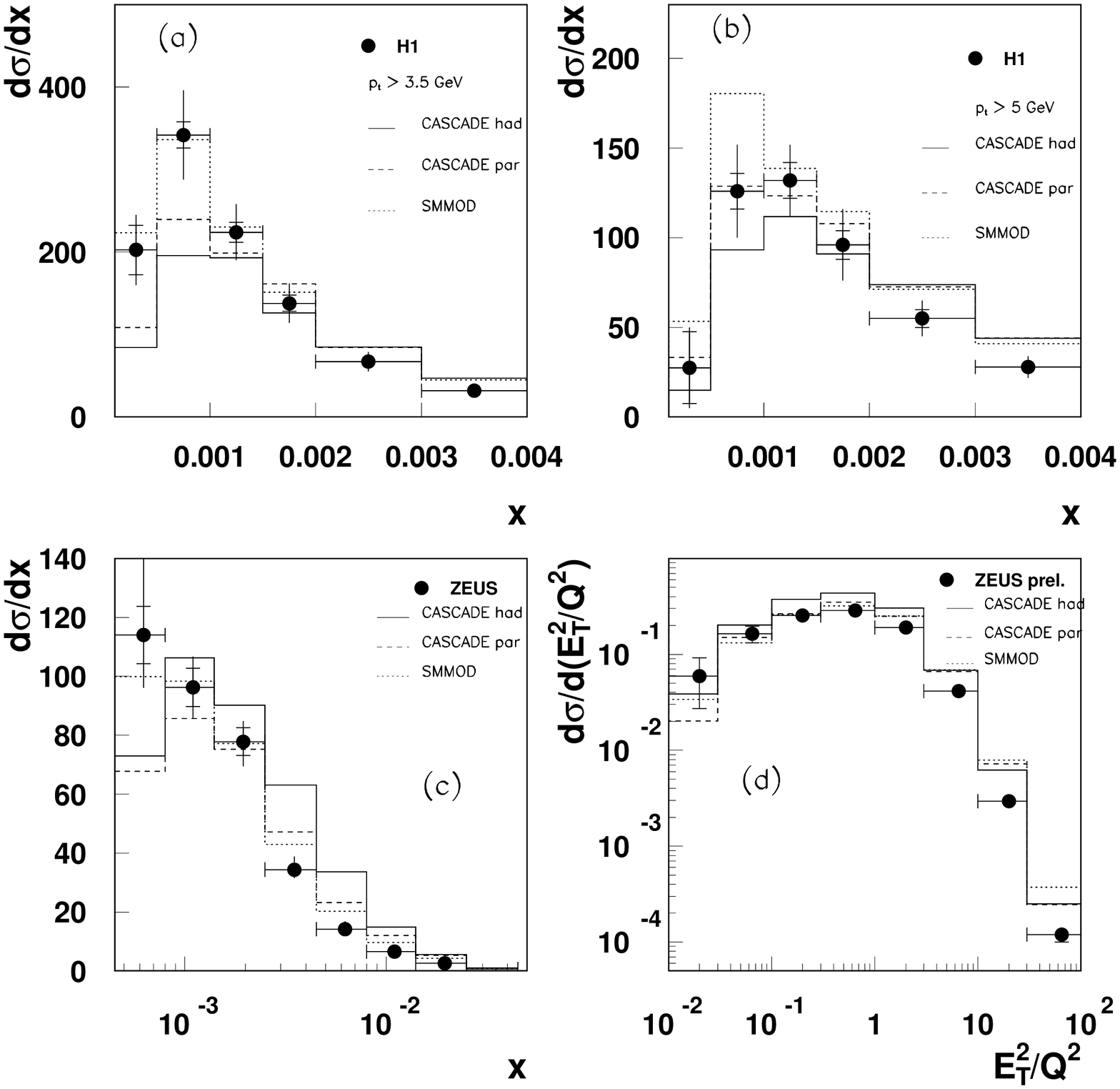,
width=18cm,height=17cm}
\caption{{\it
Comparison of the cross section for forward jet production
obtained from the backward evolution 
Monte Carlo  \CASCADE~(dashed line) at parton level
with \SMMOD~ (dotted line) at parton level.
Also shown is the prediction from \CASCADE~at hadron level (solid line).
$a.-c.$ The cross section for forward jet production as a function of $x$, 
for different cuts in $p_t$ compared to H1 data~\protect\cite{H1_fjets_data}
($a.-b.$) and 
compared to ZEUS data~\protect\cite{ZEUS_fjets_data} ($c.$).
$d.$ The cross section for forward jet production as a function of $E^2_T/Q^2$
compared to \protect\cite{ZEUS_fjets_pt2/q2}.
  }}\label{fwdjet_comp}
\end{figure}
In Fig.~\ref{fwdjet_comp} the prediction of the
cross section for forward jet production is shown. 
The parton level results from \CASCADE~ 
are similar to the ones obtained from \SMMOD, keeping in mind the differences
already seen in $F_2$.
 Also shown are the hadron level
results from \CASCADE, which differ from the parton level results by $\sim
$ 20 \%, which is known as the hadron level correction.
\par
\begin{figure}[htb]
  \vspace*{2mm}
\epsfig{figure=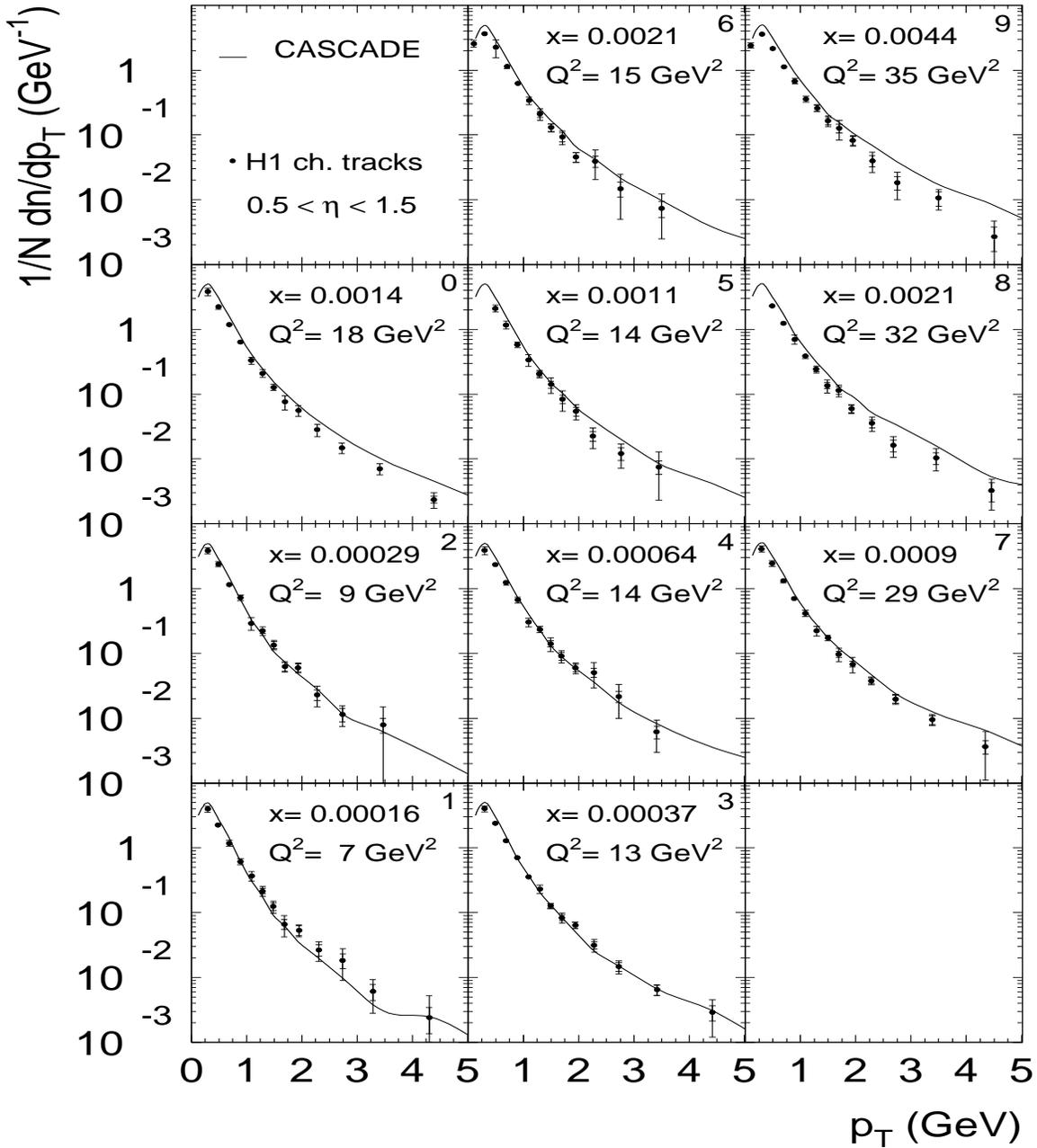,
width=17cm,height=19cm}
\caption{{\it
The transverse momentum distribution of particles in different bins of rapidity.
The prediction of the Monte Carlo \CASCADE~  at hadron level is
 compared to the measurement of H1~\protect\cite{H1_ptspectra_data}
  }}\label{particle_spectra}
\end{figure}

In Fig.~\ref{particle_spectra} the transverse momentum distribution for single
particles are shown in different bins of the rapidity. We observe a rather good
description of the experimental measurements.
\section{Conclusion}
In this article I have shown, that the CCFM evolution equation
with a proper treatment of the non-Sudakov form factor, 
gives a good description of the structure function $F_2$ and at the same time
also describes the cross section for forward jet production. 
This results were obtained from \SMMOD,  a 
modified version of the forward evolution Monte Carlo program
\SMALLX~\cite{\SMALLXC}, where the important change
was the modified treatment of the non-Sudakov form factor. Whereas for
$F_2$ equally good descriptions are obtained with and without the
``consistency constraint", the description of forward jet production improves if
no ``consistency constraint" is applied. 
\par
To fully simulate the hadronic final state of small $x$ processes including
parton emissions according to the CCFM evolution equation, a new hadron level
Monte Carlo event generator, \CASCADE, using a backward parton cascade
approach, has been developed. The unintegrated gluon density  as a
function of $x$, $k_t^2$ and $p^2$ was obtained from \SMMOD.
\par
I have shown that the parton level results obtained from the forward evolution
program \SMMOD~ agree rather well with those from the backward evolution program
\CASCADE. The hadron level results obtained from \CASCADE~ show rather nice
agreement with measurements of the HERA experiments, that are sensitive to small
$x$ parton dynamics. With standard hadron level Monte Carlo programs using DGLAP
type parton evolution and direct photon interactions only, a similar level
 of agreement could not be achieved.
\section{Acknowledgments}
I am very grateful to B. Webber for providing me with the \SMALLX~ code,
which was
the basis of the studies presented here, and his
patience to answer many questions about \SMALLX.  
I am also very grateful to 
G. Salam, H. Kharraziha, L. L\"onnblad for all their criticism and their 
clever and useful ideas on CCFM and the backward evolution. I have
learned very much not only about CCFM from discussions with B. Andersson and 
G. Gustafson. I am also grateful to R. Peschanski and S. Munier for many
discussions about CCFM, BFKL and the ``consistency constraint" and the good
 times we always have in Paris. I also want to thank A.D. Martin and 
 J. Kwiecinski for their explanation of the modified non - Sudakov form factor.
I learned much from a continuous dialogue with G. Ingelman and T.
Sj\"ostrand.

\end{document}